\documentclass{iaus}
\usepackage{graphicx}

\def\lya{\ifmmode {\rm Ly}\alpha~ \else Ly$\alpha$~\fi}
\def\lyb{\ifmmode {\rm Ly}\beta~ \else Ly$\beta$~\fi}
\def\lyg{\ifmmode {\rm Ly}\gamma~ \else Ly$\gamma$~\fi}
\def\civ{\ifmmode {\rm C}\,{\sc iv}~ \else C\,{\sc iv}~\fi}
\def\cvi{\ifmmode {\rm C}\,{\sc vi}~ \else C\,{\sc vi}~\fi}
\def\cvin{\ifmmode {\rm C}\,{\sc vi} \else C\,{\sc vi}\fi}
\def\hi{H\,{\sc i}~}

\def\nvin{N\,{\sc vi}}
\def\nvii{N\,{\sc vii}~}

\def\ovi{{{\rm O}\,{\sc vi}~}}
\def\ovii{{{\rm O}\,{\sc vii}~}}
\def\oviii{{{\rm O}\,{\sc viii}~}}
\def\ovin{{{\rm O}\,{\sc vi}}}
\def\oviin{{{\rm O}\,{\sc vii}}}
\def\oviiin{{{\rm O}\,{\sc viii}}}
\def\neix{{{\rm Ne}\,{\sc ix}~}}

\def\gax    {${_>\atop^{\sim}}$}

\def\chandra {{\it Chandra}~}
\def\xmm {{\it XMM-Newton}~}

\title[Lost Baryons at Low Redshift] 
{Lost Baryons at Low Redshift}

\author[Smita Mathur]   
{Smita Mathur$^1$, Fabrizio Nicastro$^2$, \and Rik Willimas$^3$ }

\affiliation{$1$ Astronomy Department, The Ohio State University,
Columbus, OH 43210\break email: smita@astronomy.ohio-state.edu
\\[\affilskip] $2$ OAR-INAF, Rome, Italy\\[\affilskip] $3$ Leiden
University, Netherlands }

\pubyear{2007}
\volume{244}  
\pagerange{1--9}
\date{?? and in revised form ??}
\jname{Dark Galaxies and Lost Baryons}
\editors{J. Davis \& M. Disney, eds.}
\begin{document}

\maketitle

\begin{abstract}
We review our attempts to discover lost baryons at low redshift with
``X-ray forest'' of absorption lines from the warm-hot intergalactic
medium. We discuss the best evidence to date along the Mrk 421
sightline. We then discuss the missing baryons in the Local Group and
the significance of the z=0 absorption systems in X-ray spectra. We
argue that the debate over the Galactic vs. extragalactic origin of
the z=0 systems is premature as these systems likely contain both
components. Observations with next generation X-ray missions such as
Constellation-X and XEUS will be crucial to map out the warm-hot
intergalactic medium.

\keywords{atomic processes, Galaxy: formation, intergalactic medium,
quasars: absorption lines, Local Group, cosmology: observations,
X-rays: galaxies}

\end{abstract}

\firstsection 
\section{Introduction: The Missing Baryons}

Given the title of this symposium, it should not come as a surprise to
know that most of the baryons at low redshift are missing; here is how
we know. The theory of big-bang nucleosynthesis, together with the
observations of the deuterium abundance implies that the
ratio of baryon density to the critical density of the Universe is about
4\%: $\Omega_b \approx 0.04h_{70}^{-2}$ (Burles \& Tytler 1998). The
concordance cosmology with the new WMAP observations also predicts a similar
number for $\Omega_b$ (Bennett et al. 2003). This theoretical
expectation is consistent with the observations of Lyman $\alpha$ forest
at high redshift. At low redshift, however, the Lyman $\alpha$ forest
thins out, so the baryons clearly do not stay in the warm, photoionized
intergalactic medium at low redshift, as they did at high
redshift. Perhaps, as the Universe evolved, and more and more galaxies
formed, the baryons were accreted onto the virialized systems. The
baryon census at low redshift, however, proved it not to be the
case. All the stars and gas in galaxies account for only 10\% of all the
baryons (Fukugita, Hogan \& Peebles 1998). Hot gas in clusters of
galaxies makes a similar contribution to $\Omega_b$. Cold gas at low
redshift also adds to the baryon budget, but most of the baryons were
found to be lost. This is the so called missing baryon problem.

Cosmological hydrodynamic simulations found that as the Universe
evolves, the gas in the intergalactic medium gets shock heated to high
temperatures ($10^5$--$10^7$K) and forms a web like large-scale
structure of filaments and sheets. Galaxies and clusters of galaxies
form in dense knots in the filamentary structure, but even at low
redshift most of the baryons still reside in the diffuse intergalactic
medium (IGM). The temperature of the IGM is too high to keep hydrogen
neutral, which is why Ly$\alpha$ forest thins out. Given the predicted
temperature range, the low redshift IGM was called warm-hot
intergalactic medium, or WHIM, by Cen \& Ostriker (1999), which is
hotter than the warm IGM at high redshift, but cooler than the hot gas
in clusters of galaxies.

Even though most of the hydrogen in the WHIM is ionized, most of the
metals are not fully ionized. Thus, in spectra of distant quasars,
the WHIM may imprint its signature by absorption lines of highly
ionized metals. Oxygen being one of the most abundant metals,
absorption lines of \ovii are expected to be strongest (see 
fig.1, adapted from Mathur et al. 2003. \ovii is the
dominant ionization state of oxygen in the WHIM temperature range).

\begin{figure}
\begin{center}
 \includegraphics[height=3in,width=3in]{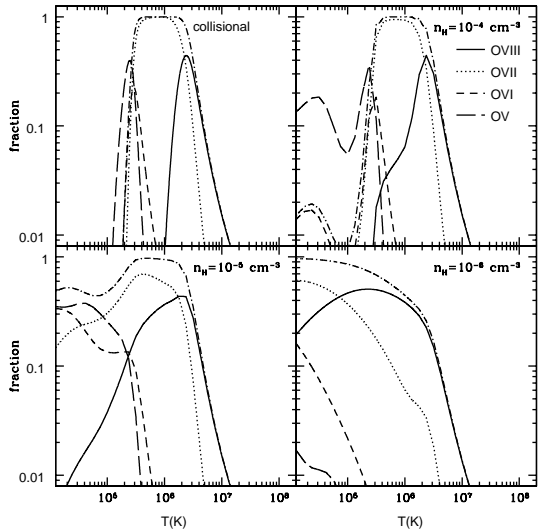}
\end{center}
  \caption{The fraction of oxygen in different ionization states
  vs. temperature. The top left panel is for the pure collisional
  ionization while photoionization from metagalactic UV and X-ray
  background is included in the other panels in which the density is
  $10^{-4}$, $10^{-5}$, and $10^{-6} cm^{-3}$ from top right to bottom
  right. }\label{fig:wave}
\end{figure}


\section{The Quest for WHIM}\label{sec:xray}

The resonance transitions of the dominant ions fall in the X-ray
band-pass. The strongest line of \ovii, for example, is at
21.602\AA. One thus expects to detect an ``X-ray forest'' of
absorption lines from hydrogen-like and helium-like ions in X-ray
spectra of distant quasars. Such an experiment, however, could not be
performed before the launch of \chandra and \xmm observatories, which,
for the first time, carry gratings on board allowing high resolution
spectroscopy. Even then, this is a very difficult experiment because
these X-ray telescopes are small and given how faint the quasars are,
the S/N of spectra becomes a limiting factor in detecting weak and
narrow absorption lines due to WHIM. For this reason we came up with a
strategy of observing blazars in their high state, which would provide
temporarily bright light beacons.

\subsection{Chandra, FUSE and HST observations of Mrk 421}

One such blazar is Mrk 421 which we monitored with the Rossi X-ray
Timing Explorer (RXTE). When the Mrk 421 flux was almost two orders of
magnitude larger than its quiescent flux, we triggered a
target-of-opportunity (TOO) observation with \chandra low energy
transmission grating (LETG). Thus we obtained the highest S/N grating
spectrum among all targets ever observed with \chandra (see figure 2,
adapted from Nicastro et al. 2005a,b). The black dots with error bars are
the data and the blue line running through the spectrum is the Mrk 421
continuum folded with the instrumental response. The red lines mark
the absorption lines in the spectrum which we identify with
transitions of several ions at redshifts zero, 0.011 and 0.027. The
ions are labeled and the three redshifts are marked with vertical
bars. The middle left panel shows the strong transition of \ovii
K$\alpha$ which is clearly detected at all the three redshifts. The
strongest line is at z=0, and we will come to this in a later section.

\begin{figure}
\begin{center}
 \includegraphics[height=5in]{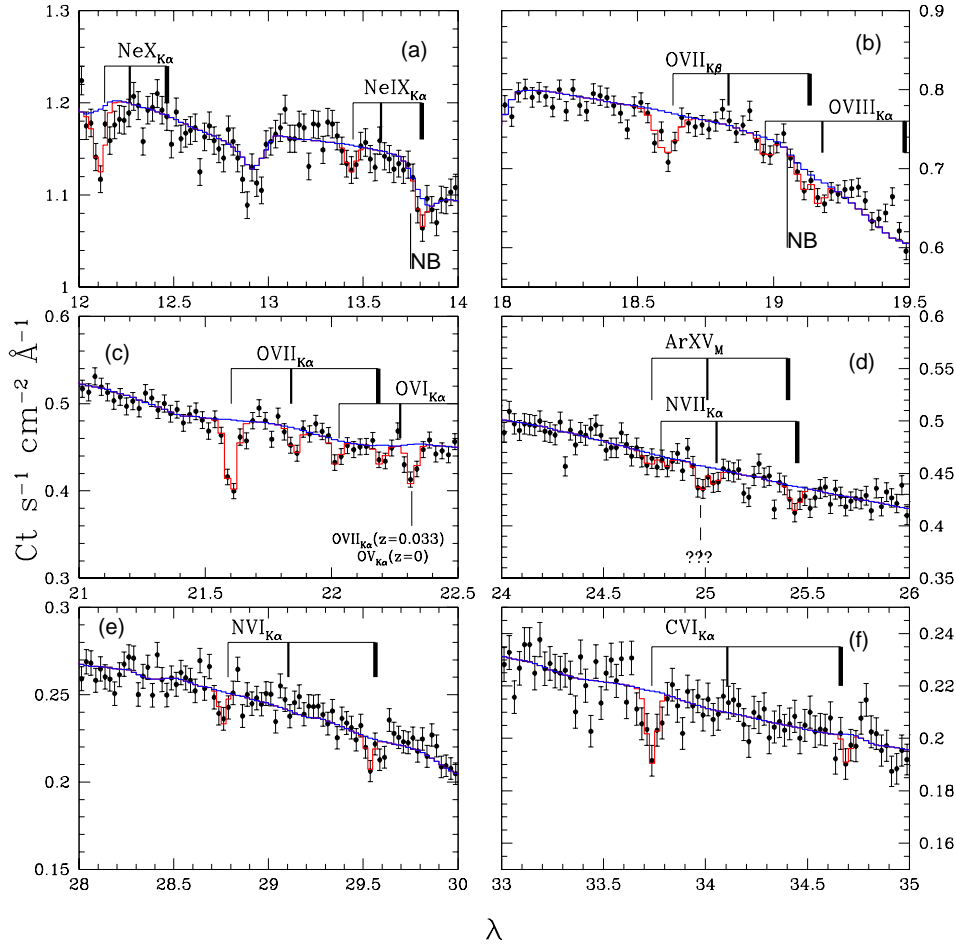}
\end{center}
  \caption{The \chandra LETG spectrum of Mrk 421. Note the detection of
  absorption lines from various H-like and He-line ions at redshift zero
  and at z=0.011 and z=0.027 marked with vertical bars.} 
  \label{fig:chandra}
\end{figure}

We also observed Mrk 421 with FUSE to look for any possible absorption
lines of \ovin. The FUSE spectrum, shown in figure 3 (adapted from
Nicastro et al. 2005b), is rich with absorption lines from the
interstellar medium of the Galaxy. However, \ovi absorption lines from
the two intervening X-ray absorption systems are not detected.

\begin{figure}
\begin{center}
 \vspace*{-1in}
 \includegraphics[width=4.5in]{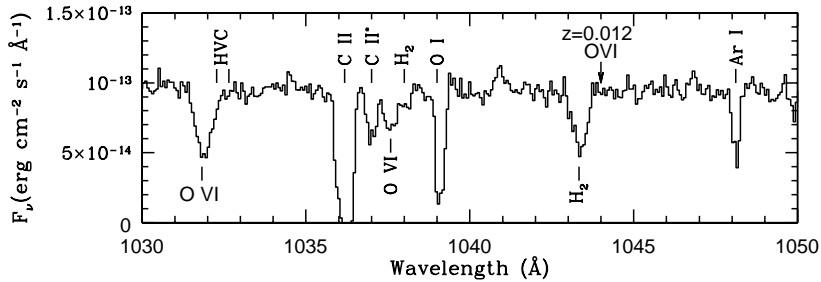}
 \vspace*{-1in}
\end{center}
  \caption{The FUSE spectrum of Mrk 421. No \ovi absorption lines are
  detected at the redshifts of intervening X-ray absorption
  systems.} \label{fig:fuse}
\end{figure}

Mrk 421 was also observed with HST and a portion of the STIS spectrum
is shown in figure 4 (adapted from Nicastro et al. 2005b). Again, we do
not detect HI Ly$\alpha$ absorption from the two intervening X-ray
systems.

To summarize, we detect two intervening absorption systems toward Mrk
421. The z=0.011 system is detected through strong absorption lines of
\ovii and \nvii, though \ovi and \hi are not detected. The z=0.027
system is detected through strong absorption lines of \oviin, \cvin,
\nvin, \nvii and less significant lines of \neix and \ovii
K$\beta$. \ovi is not detected in this systems as well and \hi
detection is also not strong, if at all present.

With the column density of detected ions and the upper limits obtained
from non-detections we determine the physical properties of the
absorbing gas. The method can be understood from figure 1 in which
fractional ionization of oxygen is plotted as a function of
temperature. The top left panel corresponds to pure collisional
ionization; we see that the fraction of oxygen in a given ionization
state is a strong function of temperature and higher ionization states
reach their peak fraction at higher temperatures. Thus, the column
density ratios of two ions depend upon temperature.  

In the other three panels of figure 1, photoionization from metagalactic
UV and X-ray backgrounds is also included in calculating ionization
fractions. As expected, the shapes of the curves change when
photoionization is added. The three panels correspond to three different
densities of the IGM, which are affected by photoionization to different
degrees. Thus, the ionic ratios become not only dependent on the
temperature, but also on the density of the IGM. In particular, we find
that the \oviin/\ovi ratio is a sensitive probe of the temperature of
the IGM, while the \oviiin/\ovi ratio measures the density (Mathur et
al. 2003). This can be understood intuitively; if a system contains both
\oviii and \ovi with measurable amounts, then photoionization must play
a role, and lower the density, higher the effect of the photoionization.

We can then make figures similar to figure 1 for other elements and
generate expected values of column density ratios for a variety of
ions. We can then compare our observed line ratios with the
theoretical calculations to constrain the properties of the
intervening systems. One such diagnostic plot is shown in figure 5
(adapted from Nicastro et al. 2005b) in which various ion ratios are
plotted vs. temperature. The thin curves represent the theoretical
 models while the thick curves represent observed values within
1$\sigma$ errors. All the observations overlap in the temperature
range of $ 0.4$--$3.3 \times 10^6$K.

Strong detection of oxygen and non-detection of \hi allow us to put a
constraint on the abundance ratio to be [O/H]\gax $-1.46$, a reasonable
value for the WHIM. The total equivalent column density is
N$_H=0.3$--$7.3 \times 10^{19} \times 10^{[O/H]_{-1}}$ cm$^{-2}$. We
have no direct constraint on the density, but if it of the order of
$10^{-5}$cm$^{-3}$, then the inferred path length is of the order of a
Mpc: D=$0.1$--$2.4 \times 10^{[O/H]_{-1}} (n_{-5})^{-1}$. {\it Thus all
the inferred parameters of the absorption systems are consistent with
those of the WHIM}.

We have detected two intervening \ovii systems in the pathlength out to
Mrk 421. This is just one sightline and there are only two systems in
it, so we are in the regime of low number statistics, but nonetheless we
can calculate the number density of \ovii filaments, albeit with a
large error bar. In figure 6 we compare this with the theoretical curve
and again find consistency. The theoretical curve in new simulations
(Cen \& Fang 2006) has lower normalization, but the Mrk 421 data point
is still consistent with models within 2$\sigma$.

\begin{figure}
\begin{center}
 \includegraphics[height=2in]{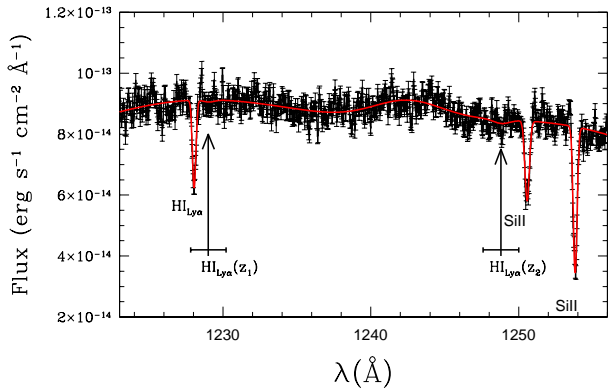}
\end{center}
  \caption{The HST spectrum of Mrk 421 showing no HI absorption from
  the two intervening X-ray systems} \label{fig:hst}
\end{figure}

\subsection{Lost is Found?}

Given the consistency of all the observed parameters with the
theoretical expectations for the WHIM, it is reasonable to assume that the
observed systems are indeed WHIM filaments. We can then calculate the
baryon density associated with these systems and compare it with the
missing baryons. We find

$$ \Omega_b = 0.032^{+0.042}_{-0.021}\times 10^{-[O/H]_{-1}} $$

\noindent
 ~\\
consistent with the missing $\Omega_b$. We are, however, cautious about
claiming that the lost baryons have been found because, once again, the
errors on our number are large due to small number statistics. It is
therefore imperative to observe more sightlines and detect more WHIM
systems. We observed a sightline toward 1ES$1028+511$ with \xmm and have
tentative detections of a couple of more systems. 1ES$1028+511$ is at a
higher redshift (z=0.361) than Mrk 421 (z=0.03), resulting in a smaller
number density of \ovii systems along this sightline. Note, however,
that the column density threshold in the \xmm spectrum of 1ES$1028+511$
is much larger and the observed number density is still consistent with
the theoretical curve in figure 6. If these are confirmed, then our
solitary point on the curve in figure 6 will have company and the errors
on $\Omega_b$ will be lowered.

\begin{figure}
\begin{center}
 \includegraphics[height=3in]{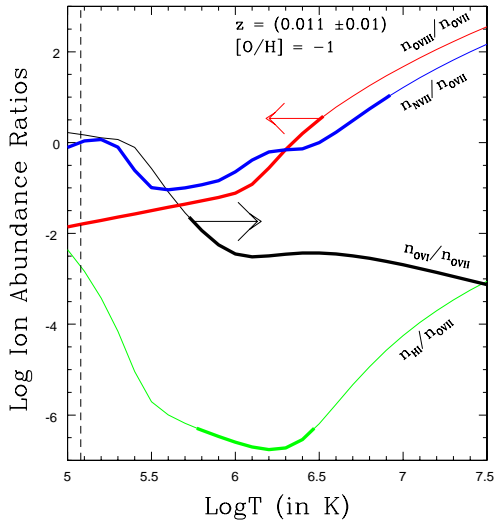}
\end{center}
  \caption{Constraints on the z=0.011 system from observed line
  ratios. The thin lines are the theoretical models while the
  thick portions on the lines mark observed values with errors. The
  temperature range where all observations overlap is around
  $10^6$K.}\label{fig:wave}
\end{figure}

\section{The z=0 Absorption Systems}\label{sec:zzero}

In high quality spectra of several extragalactic sources, absorption
lines of \ovii K$\alpha$ are detected at redshift zero (see figure 2 for
the z=0 \ovii line in the Mrk 421 sightline). The origin of such
absorption systems is debated in the literature; is it from the ISM or
the halo of our own galaxy or is it from larger scale structures such as
the Local Group? It should be noted that \ovi absorption is also
ubiquitous in extragalactic sightlines observed by FUSE. One normally
detects a broad component of \ovi at redshift zero from the thick disk
of our own Galaxy and often an additional component at a velocity offset
of more than 100 km/s from the so called high velocity cloud. The
spectral resolution in the X-ray band is not high enough to resolve the
components seen in the FUV. It is not immediately obvious, therefore,
whether the \ovii line is related to any of the \ovi components. The
answer to this question is related to the origin of \ovii systems.

\subsection{Galactic or Local Group?}

This has become such an important question because we would like to
know the associated mass. If the z=0 systems are from the local large
scale structure, then the associated mass can be huge. It may even
account for the missing baryonic mass in the Local Group. On the other
hand, if the z=0 \ovii systems are from the halo of our Galaxy, then
they may carry insignificant amount of mass.

Theoretically, some models of galaxy formation and evolution predict hot
halos around galaxies. Similarly, as mentioned in \S 1, in the
cosmological simulations galaxies form in dense knots in IGM filaments,
so we expect to find a local WHIM filament(s) in which our Galaxy is
embedded. This expectation was further confirmed with constrained
simulations of the local volume (Kravtsov et al. 2002) which show large
reservoirs of \ovii gas in extended structures around the Galaxy. Thus,
understanding the origin of \ovii systems will help test the models of
galaxy formation and/or large scale structure.

We have been investigating this question systematically (Nicastro et
al. 2002, Williams et al. 2005, 2006b, 2007) by modeling the observed
\ovii and \ovi systems. We find interesting differences among different
sightlines implying complexity in the structure surrounding the
Galaxy. We could not determine the origin of the \ovii systems; neither
Galactic nor extragalactic origins could be ruled out. One conclusion
that we could reach is that the \ovii absorption lines do not arise in
the gas producing the \ovi systems. This is most directly seen in the
Mrk 279 sightline (figure 7, adapted from Williams et al. 2006) where we
found a velocity offset between the \ovii line and the \ovi HVC.

To summarize, \ovii absorption is not associated with any single
\ovi component. Neither Galactic or extragalactic origin of the \ovii
systems can be ruled out. Most likely, a variety of phenomena are
responsible for the \ovii systems as indicated by the differences in
their ``b'' values (velocity dispersion parameters). Better, higher
resolution, data are crucial in understanding the origin of \ovii
systems.

\begin{figure}
\begin{center}
 \includegraphics[height=3in]{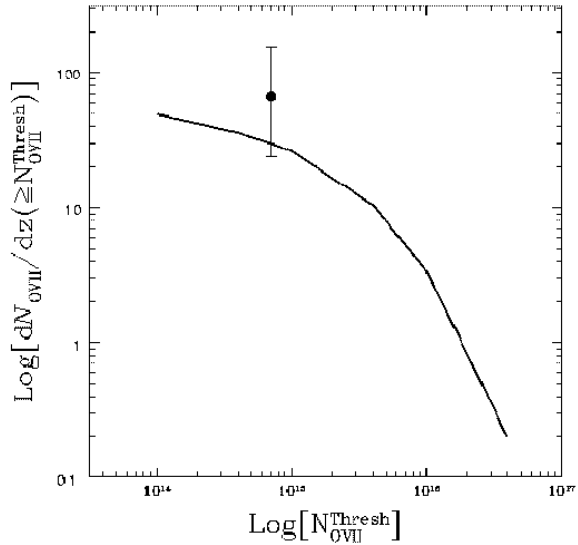}
\end{center}
  \caption{The cumulative number density of \ovii filaments due to WHIM
  is plotted as a function of observational column density
  threshold. The theoretical curve is from Fang et al. 2002 and the data
  point with error bars is from our observations. }\label{fig:wave}
\end{figure}
%

%

\section{Conclusions \& Future Prospects}\label{sec:concl}

As mentioned by C. Frenk (these proceedings) the claims of the discovery
of WHIM have been controversial. The best observed evidence, nonetheless,
is from the Mrk 421 sightline in which the chance probability of detecting
the observed absorption features by statistical fluke is less than
1\%. Detections with higher significance are desirable, of
course. Given the importance of the WHIM discovery, we need to observe
more sigh-lines and solidly detect at least a few more X-ray forest
systems before counting the baryons. The community needs to support
observing programs with these goals. It is worth noting at this point
that because of the viewing angle constraints on \chandra
observations, TOO observations of blazars in high state is not a
viable strategy any more.

Even if we succeed in our effort to detect WHIM with \chandra or \xmm,
we will be only observing the tip of the iceberg (see figure 8). To map
out the WHIM filaments at low redshift, by going down to the column
density thresholds much more representative of the large scale
structure, Constellation-X and XEUS will be essential. These are future
missions of NASA and ESA respectively; mapping the low redshift WHIM is
one of their major science drivers. We need to obtain a definite proof
of concept with \chandra and \xmm before these mission concepts are
finalized.

The science of the z=0 \ovii systems is equally interesting which
bears directly on the models of galaxy formation and/or models of
large scale structures. The spectral resolution offered by \chandra
and \xmm gratings is just not good enough to understand the origin of
these systems. The debate over the Galactic vs. extragalactic origin
of \ovii systems is premature because we do not have data quality to
resolve the issue. Given what we understand from other poor groups
similar to the Local Group, the \ovii systems must contain both of
these components. We propose that the \ovii lines contain two (or
more) components (Mathur et al. 2007) which will be resolved in higher
resolution spectra. Future higher resolution observations are
crucial to test this prediction and to make significant progress in
this field.

\begin{figure}
\begin{center}
 \includegraphics[height=2in,width=2in]{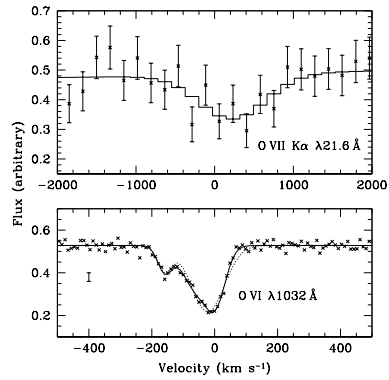}
\end{center}
  \caption{The X-ray and UV z=0 systems toward Mrk 279 show velocity
  offset}\label{fig:wave}
\end{figure}

\begin{acknowledgments}
This work is supported in part by the Chandra grant AR5-6017X.
\end{acknowledgments}

\begin{figure}
\begin{center}
 \includegraphics[height=3in]{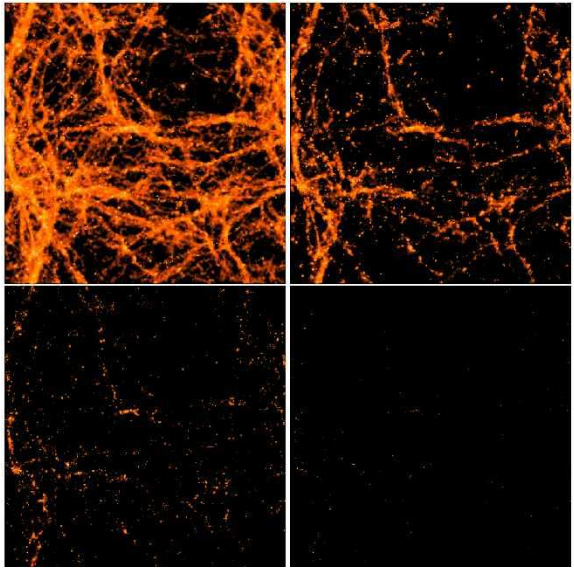}
\end{center}
  \caption{The \ovii column density map from the cosmological
  hydrodynamic simulation of the WHIM (from Chen et al. 2003). The top
  left panel has a column density threshold of $10^{14}$ cm$^{-2}$. The
  other panels from top right to bottom right correspond roughly to
  detection thresholds of XEUS, {\it Constellation-X}, and \chandra/\xmm
  respectively (assuming metallicity of a tenth solar).}\label{fig:wave}
\end{figure}
%

\begin{discussion} 

\discuss{Bomans}{The
     intervening absorption lines in the \chandra spectrum of Mrk 421
     are not detected with \xmm. Could you comment on that?}

\discuss{Mathur}{Because of the higher S/N in \xmm spectrum, it was
     natural to expect that the \chandra lines will be detected in the
     \xmm spectrum. However, S/N is not the only factor in deciding the
     line detection threshold. The shape of the line response function
     makes a big difference, just the same way a point spread function
     affects the point source detection in imaging observations. The
     line spread function of \xmm gratings has wings, increasing the
     equivalent width limit of line detections. There are a couple of
     more factors, which we have detailed in Williams et al. 2006a. The
     bottom line is that \xmm non-detections are consistent with
     \chandra detections. Later, Rasmussen et al. re-analyzed the \xmm
     data. Their upper limits are again consistent with \chandra
     detections, given the values in the paper, even though the text of
     their paper claims inconsistency!} 

\discuss{Schaye}{We do not know the metallicity of the WHIM, so we can
 never know whether the missing baryons are found. It would be much more
 fruitful to assume that the missing baryons are found and use that to
 constrain the UV and X-ray background} 

\discuss{Mathur}{Good point! 
 However, the errors on the observations at present are too large to
 constrain the background parameters. Secondly, if we actually detect
 hydrogen with UV spectroscopy from the WHIM systems, then we {\it can}
 measure metallicity.} 

\end{discussion} 


\begin{thebibliography}{}

\bibitem[]{}
     {Bennett, C.L. et al.} 2003,
      \textit{ApJS} 148, 1
\bibitem[]{}
     {Burles, S. \& Tytler, D.} 1998,
     \textit{Space Science Reviews} 84, 65
\bibitem[]{}
     {Cen, R. \& Ostriker, J.} 1999,
     \textit{ApJ} 514, 1
\bibitem[]{}
     {Cen, R. \& Fang, T.} 2006,
     \textit{ApJ}  650, 573
\bibitem[]{}
     {Chen, X., Weinberg, D., Katz, N. \& Dave, R.} 2003,
     \textit{ApJ} 594, 42
\bibitem[]{}
     {Fang, T., Bryan, G.L. \& Canizares, C.} 2002,
     \textit{ApJ} 564, 604
\bibitem[]{}
     {Fukugita, M., Hogan, C. \& Peebles, P.J.E.} 1998,
     \textit{ApJ} 503, 518
\bibitem[]{}
     {Kravtsov, A., Klypin, A. \& Hoffman, Y.} 2002
     \textit{ApJ} 571, 563
\bibitem[]{}
     {Mathur, S., Weinberg, D. \& Chen, X.} 2003,
     \textit{ApJ} 582, 82
\bibitem[]{}
     {Mathur, S., Williams, R. \& Nicastro, F.} 2007,
     \textit{in preparation} 
\bibitem[]{}
     {Nicastro, F. et al.} 2002
     \textit{ApJ} 573, 157
\bibitem[]{}
     {Nicastro, F. et al.} 2005a
     \textit{Nature} 433, 495
\bibitem[]{}
     {Nicastro, F. et al.} 2005b
     \textit{ApJ} 629, 700
\bibitem[]{}
     {Williams, R.J., et al.} 2005
     \textit{ApJ} 631, 856
\bibitem[]{}
     {Williams, R.J., Mathur, S. \& Nicastro, F.} 2006a
     \textit{ApJL} 642, 95
\bibitem[]{}
     {Williams, R.J., Mathur, S. \& Nicastro, F.} 2006b
     \textit{ApJ} 645, 179
\bibitem[]{}
     {Williams, R.J., et al.} 2007 \textit{ApJ} in press.
\bibitem[]{}
     {}  \textit{}
\bibitem[]{}
     {}  \textit{}
\bibitem[]{}
     {}  \textit{}
\bibitem[]{}
     {}  \textit{}    
\end{thebibliography}
\end{document}